\documentclass[12pt]{article}
\parskip=\medskipamount

\usepackage{amssymb,amsfonts,amsmath}
 \usepackage{bbm}

\title{A GEOMETRIC MODEL \\
OF ARBITRARY SPIN MASSIVE PARTICLE}
\author{S.M. Kuzenko, S.L. Lyakhovich\footnote{e-mail: SLL@fftgu.tomsk.su}, and
A.Yu. Segal}
\date{Department of Theoretical Physics, \\
Tomsk State University, 634050 Tomsk, Russia}

\begin{document}
\stepcounter{footnote}

\maketitle

\begin{abstract}

A new model of relativistic massive particle with arbitrary spin
(($m,s$)-particle) is suggested. Configuration space of the model is a
product of Minkowski space and two-dimensional sphere, ${\cal M}^6 =
{\Bbb R}^{3,1} \times S^2$. The system describes Zitterbewegung at the
classical level. Together with explicitly realized Poincar\'e
symmetry, the action functional turns out to be invariant under two
types of gauge transformations having their origin in the presence of
two Abelian first-class constraints in the Hamilton formalism. These
constraints correspond to strong conservation for the phase-space
counterparts of the Casimir operators of the Poincar\'e group.
Canonical quantization of the model leads to equations on the wave
functions which prove to be equivalent to the relativistic wave
equations for the massive spin-$s$ field.
\end{abstract}
\newpage
\addtocounter{footnote}{-2}

\section{Introduction}

The progress in the string theories stimulated an intensive development
of another, more traditional, direction in high-energy physics concerning
the construction of mechanical models for point-like spinning
particles and the study of their dynamical properties. First of all,
the interest to spinning (super)particles is caused by the fact that
the spectra of free (super)strings contain particle-like excitations of
higher spins. One more important point is that the superparticle models
possess a lot of features being highly similar to those arising in
superstrings (infinite reducibility of gauge algebra, nontrivial mixing
of first- and second-class constraints), what makes it possible to
exploit these models for polishing out the string quantization methods.
Finally, the systematization and study of spinning particles models
are necessary for accomplishing the standard interpretation
of quantum field theory as a second quantized theory of relativistic
particles.

Nowadays it is too early to speak about any systematization of
particle models, since new and new models appear permanently. In this
regard, it is worth mentioning the models desribing spin-1/2
particle$^{1-4}$, (extended) superpart\-ic\-les$^{5-9}$, higher spin
particles$^{10-13}$.

In most of the spinning particle models, anticommuting variables
are incorporated into configuration space to describe the spin
degrees of freedom. Such a pseudoclassical approach is known to be well
adapted for constructing the quantum theory. It makes, however,
impossible to provide an elegant geometric framework for the
classical evolution of spinning particles, which could be considered as
a natural generalization of that existing for the spinless particle.
Supposing the existence of such a geometric formulation for spinning
particles, one can expect that the corresponding configuration space
should have sufficiently low dimension to describe only the space-time
evolution and the spin dynamics. Equivalently, every physical
observable commuting with the first-class constraints must be a function of
the Poincar\'e generators on the phase space. This implies in turn
that the spin variables have to transform nonlinearly with respect to the
Lorentz group.

The Lorentz group $SO(3,1)^\uparrow \cong SL(2,{\Bbb C})/{\Bbb Z}_2$
is usually defined to be the connected component of unit in the group
$O(3,1)$ of all linear homogeneous transformation on ${\Bbb R}^{3,1}$
preserving the metric ${\rm d}s^2 = \eta_{ab}{\rm d}x^a{\rm d}x^b$,
$\eta_{ab}$ = diag ($-+++$). On the other hand, $SL(2,{\Bbb C})/{\Bbb
Z}_2$ naturally acts on a two-dimensional sphere $S^2$, by
fractional linear transformations, and coincides with the group of all
complex automorphisms of $S^2$. This fact underlies the twistor
approach$^{11}$. It is worth noting that $S^2$ is the unique
representative in the family of $SO(3,1)^\uparrow$-transformation
spaces in the sense that it has minimally possible dimension.

Since $SO(3,1)^\uparrow$ is realized as the transformation group of the
manifolds ${\Bbb R}^{3,1}$ and $S^2$, it seems reasonable to consider
the united space ${\cal M}^6 \equiv {\Bbb R}^{3,1}\times S^2$, being
a transformation space of the Poincar\'e group (the space-time
translations are defined to act trivially on $S^2$), and to try luck in
constructing a Poincar\'e-invariant theory of particle on ${\cal
M}^6$. Obviously, such a particle model on ${\cal M}^6$ will be
equivalent to some model of spinning particle in Minkowski space
with the spin sector to be formed by the spherical modes. We propose
the requirement that the system on ${\cal M}^6$ must possess two strong
conservation laws corresponding to the mass and spin of the particle as
a basic dynamical principle which allows to choose unique action
functional.

The present paper is devoted to realization of the above program. It
will be shown that a free massive particle of arbitrary fixed spin can be
described in the framework of special model of a particle on
${\cal M}^6$. The corresponding Lagrangian has a geometric origin, and
its structure is determined by the parameters of mass $m$ and spin $s$
(($m,s$)-particle model).

The paper is organized as follows. In section 2 we develop a Lagrange
formalism for a Poincar\'e-invariant model of particle on ${\cal M}^6$.
The Lagrangian of the model involves not only the interval along the world
line in Minkowski space, but also a Poincar\'e-invariant interval on
$S^2$. The spherical metric turns out to be dynamical in the sense that
it depends explicitly on the tangent vector to the world line in
Minkowski space. Here we show that the causality principle for massive
spinning particle consists of imposing the inequality ${\dot x}^2 < 0$,
for every point of the world line, together with some relativistic
restriction on maximal velocity of particle on $S^2$. The dynamics of
($m,s$)-particle proves to be nontrivial and is similar to the known
effect of Schr\"odinger vibration (Zitterbevegung) in relativistic
quantum theory (analogous dynamical behaviour of classical spinning
particles has been observed in more early models$^{14, 15}$). In section
3 we develop a Hamilton formalism for the model of ($m,s$)-particle,
study the physical observables and integrate the equations of motion
with arbitrary Lagrange multipliers. Section 4 is devoted to the
quantization of the model of ($m,s$)-particle. In concluding section 5
we discuss our results and further perspectives. We also include into
the paper three appendices having technical character. Appendix A
contains necessary facts concerning the action of the Lorentz group on
$S^2$. Tensor fields on ${\cal M}^6$ are described in appendix B. In
appendix C we briefly present a relativistic harmonic analysis on $S^2$.

Our notations and conventions coincide mainly with those adopted in
Wess and Bagger's book$^{16}$. In particular, $\sigma$-matrices and
$\gamma$-matrices are chosen in the form
$$
({\sigma_a})_{\alpha{\dot\alpha}} = ({\mathbbm 1},{\vec\sigma}), \qquad
({{\tilde\sigma}_a})^{{\dot\alpha}\alpha} = \varepsilon^{\alpha\beta}
\varepsilon^{\dot\alpha\dot\beta}(\sigma_a)_{\beta\dot\beta}, \qquad
\gamma_a = \left( \begin{array}{cc} 0 & \sigma_a\\
{\tilde\sigma}_a & 0 \end{array} \right).
$$
But in contrast with Ref. 16, it is useful for us to number
two-component spinor indices by values 0, 1 ($\varepsilon^{01} =
\varepsilon^{{\dot 0}{\dot 1}} = 1$) and to define matrices
$\sigma_{ab}$ and ${\tilde\sigma}_{ab}$ by the rule
\begin{eqnarray*}
{(\sigma_{ab})_\alpha}^\beta = -~\frac{1}{4} \{
(\sigma_a)_{\alpha{\dot\gamma}} ({\tilde\sigma}_b)^{{\dot\gamma}\beta} - (a
\leftrightarrow b)\},\\
{({\tilde\sigma}_{ab})^{\dot\alpha}}_{\dot\beta} = -~\frac{1}{4} \{
({\tilde\sigma}_a)^{{\dot\alpha}\gamma}(\sigma_b)_{\gamma{\dot\beta}}
 - (a \leftrightarrow b)\}.
\end{eqnarray*}

\section{Model of relativistic particle on ${\cal M}^6$}

Similarly to Minkowski space ${\Bbb R}^{3,1}$, the manifold ${\cal M}^6
= {\Bbb R}^{3,1}\times S^2$ is a (homogeneous) transformation space for
the Poincar\'e group $\cal P$ and, in principle, can be chosen on the
role of an arena where dynamics of relativistic particles is
developing.

The action of the Lorentz group $SL(2,{\Bbb C})/{\Bbb Z}_2$ on
two-sphere $S^2$ is well known (see, for example, Refs. 11, 17) and
described in appendix A in the form appropriate for our purposes. It
can be continued uniquely to the action of the Poincar\'e group on
$S^2$ by attaching to each space-time translation the identity map on
$S^2$. Then, the relations (A.4) and
$$
x^{\alpha\dot\alpha} \rightarrow {x'}^{\alpha\dot\alpha} =
x^{\beta\dot\beta}{(N^{-1})_\beta}^\alpha {(\overline N^{-1})_{\dot\beta}}
^{\dot\alpha} + f^{\beta\dot\beta}
\eqno{(1)}$$
define an action of the Poincar\'e group on ${\cal M}^6$. Here $x^a =
-~\frac{1}{2} (\sigma^a)_{\alpha{\dot\alpha}} x^{\alpha{\dot\alpha}}$
are the coordinates on ${\Bbb R}^{3,1}$, $f^a = -~\frac{1}{2}
(\sigma^a)_{\alpha{\dot\alpha}} f^{\alpha{\dot\alpha}}$ are the
parameters of translations, and the Lorentz transformations are
associated in standard fashion with complex unimodular 2$\times$2
matrices (A.2).

It should be remarked that the composite structure of ${\cal M}^6$,
being the product of ${\Bbb R}^{3,1}$ and $S^2$, admits the action of
the group ${\cal P}\times{\cal P}$ on ${\cal M}^6$. However, physically
interesting theories on ${\cal M}^6$, i.e. admitting four-dimensional
interpretation, must have only the diagonal of ${\cal P}\times{\cal P}$
as a symmetry group. Before describing the fulfilment of this
requirement for a point particle on ${\cal M}^6$, we first present
some more ${\cal M}^6$-formalism.

Let us introduce a two-component object
$$
z^\alpha \equiv (z)^\alpha = (1, z), \qquad \alpha = 0,1,
\eqno{(2)}$$
constructed in terms of the complex local coordinate $z$ on $S^2$ and
connected with the coordinates $q^\alpha$ on $U_0 \subset {\Bbb
C}^2_*$ (A.1) by the rule $z^\alpha = q^\alpha /q^0$. The relations
(A.2--4) imply that $z^\alpha$ is transformed under the action of the
Lorentz group by the law
$$
z^\alpha \rightarrow {z'}^\alpha = \left(\frac{\partial z'}{\partial z}
\right)^{1/2} z^\beta {(N^{-1})_\beta}^\alpha ,
\eqno{(3)}$$
that is, simultaneously, as a tensor field on $S^2$ of type \{1/2, 0\}
(see appendix B) and a left Weyl spinor\footnote{A smooth tensor
field $\Phi^{[n/2]}$ on $S^2$ of type \{$n$/2, 0\}, which satisfies
the requirement of holomorphicity (B.6), can be represented in the form
(B.7), $F_{\alpha_1\ldots\alpha_n}$ being a Lorentz tensor of type
($n$/2, 0).}. Let $p^a$ be a time-like 4--vector,
$$
p^2 = p^ap_a < 0,
\eqno{(4)}$$
``living'' in Minkowski space. One can associate with this 4--vector a
smooth positive definite metric on $S^2$ of the form
$$
{\rm d}s^2 = \frac{4{\rm d}z{\rm d}{\bar z}}{(p^a\xi_a)^2} = 2g_{z{\bar
z}}{\rm d}z {\rm d}{\bar z},
\eqno{(5)}$$
where
$$
\xi_a = (\sigma_a)_{\alpha{\dot\alpha}} z^\alpha{\bar z}^{\dot\alpha} =
{\bar\xi}_a = (1 + z{\bar z}, z + {\bar z}, {\rm i}z -
{\rm i}{\bar z}, 1 - z{\bar z}), \quad \xi^a\xi_a = 0.
\eqno{(6)}$$
For the Lorentz transformation (A.4) we have, due to Eq. (3), the
following relation
$$
{{\rm d}z'{\rm d}{\bar z}'\over(p'_{\alpha\dot\alpha}z'^\alpha
{\bar z}'^{\dot\alpha})^2} = {{\rm d}z{\rm d}{\bar
z}\over(p_{\alpha\dot\alpha}z^\alpha {\bar z}^{\dot\alpha})^2},
$$
where
$$
{p'}_{\alpha\dot\alpha} = {N_\alpha}^\beta \left.{\overline N}_{\dot\alpha}
\right.^{\dot\beta} p_{\beta\dot\beta}
$$
is the transformation law of a 4--vector under the Lorentz group. In
this sense {\rm d}$s^2$ {\it is invariant with respect to the action of
the diagonal in ${\cal P}\times{\cal P}$ on ${\cal M}^6$}. It is worth
noting that the metric (5) is characterized by a constant positive
curvature, namely:
$$
[\nabla_z,\nabla_{\bar z}] = \frac{1}{2}(r-s)g_{z{\bar z}}R, \qquad R =
-p^2.
\eqno{(7)}$$
\setcounter{footnote}{6}
Here we have assumed the commutator to be taken on the space of tensor
fields on $S^2$ of type $\{r/2,s/2\}$\footnote{The covariant
derivative $\nabla_z$ acts on a field of type $\{r/2,s/2\}$ as
$\nabla_z = \partial_z + \frac{r}{2}\Gamma^z_{zz}$, $\Gamma^z_{zz} =
-2\partial_z \ln (p,\xi)$, where $(p,\xi) \equiv p^a\xi_a$.}. In a
rest reference system where $p^a = (p^0, 0, 0, 0)$, the above metric
coincides, up to a constant multiplier, with the standard metric on
sphere,
$$
{\rm d}s^2 \sim \frac{4{\rm d}z{\rm d}{\bar z}}{(1 + z{\bar z})^2}.
$$

Inequality (4) proves to be equivalent to the condition that Eq.
(5) defines a smooth tensor field on $S^2$ of type \{-1, -1\}. In the
case of a massive particle, there is a natural candidate on the role of
$p^a$ --- the derivative ${\dot x}^a$ along the space-time trajectory
$x^a(\tau)$ with respect to the evolution parameter $\tau$. As a
result, the requirement of Poincar\'e invariance, i.e. of invariance
with respect to the action of the diagonal in ${\cal P}\times{\cal P}$
only, for a theory of massive particle on ${\cal M}^6$ means that both
Poincar\'e scalars related to each point on the world line,
$$
{\dot x}^2 = {\dot x}^a{\dot x}_a, \qquad \frac{4{\dot z}{\dot{\bar
z}}}{({\dot x},\xi)^2},
$$
should arise in the corresponding Lagrangian.

The Lagrangian of a relativistic particle on ${\cal M}^6$ is given by
the following expression
$$
{\cal L} = \frac{1}{2e_1}\{ {\dot x}^2 - (mce_1)^2\} + \frac{1}{2e_2}
\left\{\frac{4{\dot z}{\dot{\bar z}}}{({\dot x},\xi)^2} e_1^2 +
(\Delta e_2)^2 \right\}.
\eqno{(8)}$$
Here $e_1(\tau)$ and $e_2(\tau)$ are Lagrange multipliers
(``einbeins'') associated with the particle's motion in ${\Bbb
R}^{3,1}$ and $S^2$, respectively. The Lagrangian consists of two
parts: a Minkowskian length and its spherical counterpart, the former
being well known Lagrangian of a massive spinless particle. The
parameter $\Delta$ can be interpreted as a spherical ``mass''. As will
be shown below, $\Delta$ is connected with the particle's spin. The
introduction of $e_1$ into the spherical part of $\cal L$ guarantees
the invariance of the action functional under world-line
reparametrizations looking infinitesimally like
$$
\delta_\epsilon x^a = {\dot x}^a\epsilon , \qquad \delta_\epsilon z
= \vspace{-12pt}{\dot z}\epsilon ,
$$
{~}\vspace{-12pt}\hfill (9)\\
$$
\delta_\epsilon e_i = \frac{{\rm d}}{{\rm d}\tau}(e_i\epsilon ),
\qquad i = 1,2,
$$
the parameter $\epsilon(\tau)$ being arbitrary modulo standard boundary
conditions. What is more, the action functional possesses another gauge
invariance of the form
$$
\delta_\mu x^a = p^a\mu , \qquad \delta_\mu e_1 = \dot\mu , \qquad
\delta_\mu z = \delta_\mu e_2 = 0.
\eqno{(10)}$$
Here $\mu$ denotes the gauge parameter, and $p^a$ the four-momentum of
the particle,
$$
p^a = \frac{\partial{\cal L}}{\partial{\dot x}_a} = \frac{{\dot x}^a}
{e_1} - \frac{4{\dot z}{\dot{\bar z}}}{({\dot x},\xi)^3}~ \frac{e_1^2}
{e_2}\xi^a.
\eqno{(11)}$$
The presence of two independent gauge symmetries (9) and (10) in the
theory with the Lagrangian (8) gives rise to two first-class
constraints in Hamilton formalism.

The Lagrange multipliers $e_1$ and $e_2$ can be eliminated with the aid
of their equations of motion
$$
\left( \frac{{\dot x}}{e_1} \right)^2 + m^2c^2 - \frac{8{\dot
z}{\dot{\bar z}}}{({\dot x},\xi)^2}~ \frac{e_1}{e_2} = 0;
\eqno{(12.a)}$$
$$
\frac{4{\dot z}{\dot{\bar z}}}{({\dot x},\xi)^2} \left(
\frac{e_1}{e_2} \right)^2 - \Delta^2 = 0.
\eqno{(12.b)}$$
Then ${\cal L}$ takes the form
$$
{\cal L} = -mc\sqrt{-{\dot x}^2\left( 1 - \frac{4\Delta}{m^2c^2}~
\frac{|{\dot z}|}{({\dot x},\xi)}\right)}.
\eqno{(13)}$$
For $\Delta$ = 0 this expression apparently reduces to the Lagrangian
of massive spinless particle.

It is necessary to point out that the parameter $\Delta$ is
dimensional. As is easily seen, it can not be made dimensionless by
means of redefinitions involving only another parameter of the theory,
the mass of particle, and the speed of light. Accounting, however,
the Planck constant, $\Delta$ can be written on the manner
$$
\Delta = \hbar mc\sqrt{s(s+1)},
\eqno{(14)}$$
where $s$ is a dimensionless non-negative constant. We identify $s$
with particle's spin. Classically, $s$ can take arbitrary values. In
what follows, we are going to show that consistent quantization of the
theory turns out to be possible only for (half)integer values of $s$.
As for the appearance of $\hbar$ in the classical Lagrangian (13) of
massive spinning particle (($m,s$)-particle), this seems natural from
the usual standpoint that spin is a quantum effect disappearing in the
limit $\hbar \rightarrow 0$.

Let us discuss a global structure of the space of world lines (the
space of histories) in the model of ($m,s$)-particle. Similarly to the
spinless case, the space of histories must consist only of those world
lines $\varphi^i = \{x^a(\tau ),z(\tau ),{\bar z}(\tau )\}$ consistent
with the requirement of space-time causality:
$$
{\dot x}^2 < 0, \qquad {\dot x}^0 > 0.
\eqno{(15)}$$
On the other hand, it follows from the explicit form of the Lagrangian
(13) that ${\cal L}$ is well defined under fulfilment of the following
restriction:
$$
|{\dot z}|/({\dot x},\xi ) < m^2c^2/4\Delta ;
\eqno{(16.a)}$$
here and below we set $\Delta \neq 0$. This restriction means that
($m,s$)-particle can not move with arbitrary large velocity not only in
Minkowski space, but also in the internal (spinning) space. In
addition, the requirement (16.a) should be supplemented by the
inequality
$$
{\dot z} \neq 0
\eqno{(16.b)}$$
presenting a consistency condition for the equation (12.b). At first
sight, the restrictions have a technical character. However, one
readily finds that a solution $\varphi^i_0$ of the dynamical equations
$$
\delta {\cal L}/\delta x^a = 0, \qquad \delta{\cal L}/\delta z = 0
$$
is causal (consistent with (15)) if and only if it is consistent with
(16). Hence, all admissible world lines must obey the system of
inequalities (15) and (16) which are to be understood as the full set
of causal conditions for the massive spinning particle. In virtue of
the importance of this assertion, let us give more comments.

We would like to note first that for every world line consistent with
(15) the conjugate 4--momentum of particle (11) satisfies
automatically the relations (4) and $p^0 > 0$. On the mass-shell for
$x^a$, the 4--momentum is constant, ${\dot p}^a = 0$, while the
equation (12.a) implies
$$
p^2 + m^2c^2 = 0,
\eqno{(17)}$$
as a consequence of the identity $\xi^2 = 0$. Let us choose a constant
4--vector $p^a$ satisfying (17) and the condition $p^0 > 0$. Let
us also choose arbitrary positive-definite functions $e_1(\tau )$ and
$e_2(\tau)$. Making use of the relations $({\dot x},\xi ) = e_1(p,\xi
)$ and (12.b), we rewrite the dynamical equations for the variables
$x^a$ and $z$ in the following equivalent form:
$$
{\dot x}^a = e_1p^a + \Delta^2e_2\frac{\xi^a}{(p,\xi)};
\eqno{(18.a)}$$
$$
\frac{{\rm d}^2z}{{\rm d}{\tilde\tau}^2} + \Gamma^z_{zz} \left(
\frac{{\rm d} z}{{\rm d}\tilde\tau} \right)^2 = 0, \qquad
{\tilde\tau} = \int_{\tau_0}^\tau {\rm d}s~e_2(s).
\eqno{(18.b)}$$
Here $\Gamma^z_{zz} = -2\partial_z~\ln (p,\xi )$ is the Christoffel
symbol for the metric (5). Now, it is easy to show that the system of
inequalities (15) is satisfied for a solution $\{x^a(\tau ), z(\tau
), {\bar z}(\tau )\}$ of Eqs. (18) if and only if this solution is
characterized by Eqs. (16).

We briefly run through discrete symmetries of the ($m,s$)-particle
model. One can readily find that the Lagrangian (8) is invariant under
the transformation of space inversion
$$
{x'}^0 = x^0, \qquad {\vec x}' = -{\vec x}
$$
acting on $S^2$ as the complex antiautomorphism
$$
z' = -1/{\bar z}.
\eqno{(19)}$$
Associated with the space-time inversion is the identity map on $S^2$.

An admissible way to fix the gauge freedom (9), (10) in the model of
($m,s$)-particle consists in imposing the following gauge conditions:
$$
{\dot x}^0 = c, \qquad e_2 = \frac{mc^2}{\Delta^2}\kappa = {\rm const}.
\eqno{(20)}$$
Here $\kappa$ is a dimensionless parameter. In accordance with Eqs.
(18), in the gauge chosen the dynamics on $S^2$ is specified by the
geodesic equation
$$
{\ddot z} + \Gamma^z_{zz}({\dot z})^2 = 0,
\eqno{(21)}$$
while the multiplier $e_1$ and the space components of the trajectory in
Minkowski space are determined by the trajectory on $S^2$:
$$
e_1 = \frac{c}{p^0} \left( 1 + mc\kappa \frac{1 + z{\bar z}}{(p,\xi )}
\vspace{-14pt}\right),
$$
{~~}\hfill \vspace{-14pt}(22)\\
$$
{\dot{\vec x}} = \frac{c}{p^0}\left( 1 + mc\kappa\frac{1 + z{\bar z}}
{(p,\xi )} \right){\vec p} + mc^2\kappa\frac{\vec\xi}{(p,\xi )}.
$$

For the gauge considered, the equations of motion are integrated most
simply in the case $p^a = (mc, 0, 0, 0)$. Then a particular solution
is given by the expressions
$$
z(\tau ) = \vspace{-12pt}{\rm e}^{\i\omega\tau},
$$
{~} \hfill \vspace{-12pt}(23)\\
$$
{\vec x}(\tau ) = \frac{\hbar}{mc}\sqrt{s(s+1)} \left( \begin{array}{c}
\sin~\omega\tau\\
\cos~\omega\tau\\
0 \end{array} \right) + {\vec x}(0),
$$
where the frequency of rotation $\omega$ is connected with $\kappa$ by
the rule
$$
\omega = \frac{mc^2}{\hbar\sqrt{s(s+1)}}\kappa ,
\eqno{(24)}$$
and the restrictions (16) mean
$$
0 < \omega < mc^2/\hbar\sqrt{s(s+1)}.
\eqno{(25)}$$
General solution can be restored from (23) by means of applying
$SO(3)$-rotations and inversions. As is seen from (25), the upper bound
for the frequency of rotation is determined by spin and decreases with
increasing of $s$.

Basing on the explicit form of the solution (23), it is easily to imagine
the situation having place in the case of arbitrary 4-momentum $p^a$.
Namely, in the space-time ($m,s$)-particle moves in and oscillates
around a straight line to which $p^a$ is tangent. But the amplitude of
such oscillations turns out to be of the order $\hbar$ and, hence,
disappears in the limit $\hbar \rightarrow 0$. Therefore, we come to
the conclusion that the classical dynamics in the model of
($m,s$)-particle is analogous to the phenomenon of Schr\"odinger
vibration (Zitterbevegung) in relativistic quantum theory.

To conclude this Section, we present the reformulation of the model
most closely related to Penrose's treatment of $S^2$ as the
celestial sphere$^{11}$. It turns out that the ($m,s$)-particle can be
described by the Lagrangian
$$
{\cal L} = \frac{1}{2e}\left({\dot x}^2 - (mce)^2\right) + \frac{1}{2}
\left(\frac{{\dot u}^2}{({\dot x},u)^3} e^2 + \Delta^2({\dot x},u)
\right),
\eqno{(26)}$$
where $e(\tau)$ is an einbein, and $u^a(\tau)$ is constrained to be a
light-like 4--vector $(u^0(\tau) < 0)$ at each point of the world-line.
The action functional, constructed on the base of (26), is obviously
invariant under the world-line reparametrizations
$$
\delta_\epsilon x^a = {\dot x}^a\epsilon , \qquad \delta_\epsilon u^a
= {\dot u}^a\epsilon , \qquad \delta_\epsilon e = \frac{{\rm d}}{{\rm
d}\tau}(e\epsilon ),
\eqno{(27)}$$
and possesses another local invariance
$$
\delta_\mu x^a = p^a\mu, \qquad \delta_\mu u^a = -~ \frac{u^a}{e}
{\dot\mu}, \qquad \delta_\mu e = \dot\mu ,
\eqno{(28)}$$
$p_a$ being the canonically conjugate momentum to $x^a$ with respect to
(26). The Lagrangian (26) takes the form (8) after the identification
$$
e \equiv e_1, \qquad u^a \equiv \frac{\xi^a}{({\dot x},\xi)}e_2.
\eqno{(29)}$$

\section{Hamilton formulation}

We are going now to construct constrained Hamilton formulation$^{18}$
for the model.

Starting with the Lagrangian (13), let us introduce conjugate momenta
for the variables $x^a$ and $z$ (in what follows, we set $\hbar = c =
1$):
$$
p_a = \frac{\partial{\cal L}}{\partial{\dot x}^a} = \frac{{\dot x}_a
(m^2 - 4\Delta\Psi ) + 2\Delta\Psi{\dot x}^2\xi_a/({\dot x},\xi )}
{\sqrt{-{\dot x}^2(m^2 - \vspace{-14pt}4\Delta\Psi)}},
$$
{~} \hfill \vspace{-14pt}(30)\\
$$
p_z = \frac{\partial{\cal L}}{\partial{\dot z}} = \frac{\Delta} {({\dot
x},\xi )} \sqrt{-~\frac{{\dot{\bar z}}{\dot x}^2}{{\dot z}(m^2 -
4\Delta\Psi )}}, \qquad \Psi \equiv \frac{|{\dot z}|}{({\dot x},\xi )}.
$$
Then one readily finds that the system possesses two primary
constraints:
$$
T_1 \equiv p^ap_a + m^2 = 0;
\eqno{(31.a)}$$
$$
T_2 \equiv (p^a\xi_a)^2p_zp_{\bar z} - \Delta^2 = 0, \qquad \Delta^2 =
m^2s(s+1).
\eqno{(31.b)}$$
These constraints are Abelian ones, $\{T_1,T_2\} = 0$, with
respect to the canonical Poisson bracket
$$
\{A,B\} = \frac{\partial A}{\partial x^a}~\frac{\partial B}{\partial
p_a} + \frac{\partial A}{\partial z}~\frac{\partial B}{\partial p_z} +
\frac{\partial A}{\partial{\bar z}}~\frac{\partial B}{\partial p_{\bar
z}} - (A \leftrightarrow B),
\eqno{(32)}$$
$A,B$ being scalar functions over the phase space. Because of
reparametrization invariance, the Hamiltonian vanishes,
$$
H_0 = p_a{\dot x}^a + p_z{\dot z} + p_{\dot z}{\dot{\bar z}} - {\cal L}
\equiv 0,
$$
hence there are no secondary constraints. As a result, the system
possesses two first class constraints, and the total Hamiltonian is
a linear combination of constraints,
$$
H = \frac{1}{2}\lambda_1\{p^2 + m^2\} + \frac{1}{2}\lambda_2 \{(p,\xi
)^2p_z p_{\bar z} - \Delta^2\},
\eqno{(33)}$$
with arbitrary Lagrange multipliers $\lambda_1(\tau)$ and
$\lambda_2(\tau)$. The canonical action looks like
$$
S = \int{\rm d}\tau~(p_a{\dot x}^a + p_z{\dot z} +
p_{\bar z}{\dot{\bar z}} - H).
\eqno{(34)}$$

The constraints $T_i, i = 1,2$, appearing in the model of
($m,s$)-particle, have a simple group --- theoretic interpretation. The
action of the Poincar\'e group defined on ${\cal M}^6$ can be lifted up
on the cotangent bundle $T^*({\cal M}^6)$ determining phase space of
the system. Obviously, all the Poincar\'e transformations on the phase
space are canonical. The action of ${\cal P}$ on $T^*({\cal M}^6)$
induces special representation of this group in the space of scalar
fields on $T^*({\cal M}^6)$, and the corresponding infinitesimal
Poincar\'e transformations can be written in terms of the Poisson
bracket (32) as follows
$$
\delta A = \{ A, -f^aP_a + \frac{1}{2}K^{ab}J_{ab}\}.
\eqno{(35)}$$
Here $f^a$ and $K^{ab} = -K^{ba}$ are the parameters of translations and
Lorentz transformations, respectively, and the Hamilton generators look
like
$$
P_a = p_a, \qquad J_{ab} = x_ap_b - x_bp_a + M_{ab},
\eqno{(36)}$$
where
$$
M_{ab} = -(\sigma_{ab})_{\alpha\beta} z^\alpha z^\beta p_z +
({\tilde\sigma}_{ab})_{{\dot\alpha}{\dot\beta}}{\bar z}^{\dot\alpha}
{\bar z}^{\dot\beta}p_{\bar z}.
$$
As is seen, the spinning part of the Lorentz generators is realized by
the spherical variables only. Let us construct, using the Hamilton
potentials, the classical Pauli--Lubanski vector $W^a =
\frac{1}{2}\varepsilon^{abcd}P_bJ_{cd}$ and the function of squared
spin $W^aW_a$. Now, it is a simple exercise to check the relation
$$
W^aW_a = (p^a\xi_a)^2p_zp_{\bar z}.
\eqno{(37)}$$
Therefore, Eqs. (31) mean that the functions of squared momentum and
spin (which can be treated as phase-space counterparts of the
Casimir operators of the Poincar\'e group) conserve strongly on
the constrained surface. Each weak physical observable ${\cal A}(x^a,
p_b, z, p_z, {\bar z}, p_{\bar z})$ should meet the requirements
$$
\left.\{ {\cal A}, T_i\}\right|_{T_j = 0} = 0, \qquad i,j = 1,2,
$$
what provides ${\cal A}$ to be gauge invariant on the constrained
surface. It can be shown that the most general form for physical
observables reads as
$$
{\cal A} = f(P_a,J_{bc}) + \varphi_iT_i,
\eqno{(38)}$$
$\varphi$'s being arbitrary functions on the phase space. If ${\cal A}$
is a strong physical observable, i.e. $\{ {\cal A}, T_i\} = 0$, then it
proves to depend only on the Hamilton generators (36). So every
physical observable is a function of the Hamilton generators of the
Poincar\'e group modulo constraints.

The Hamilton formulation is convenient for solving the equations
of motion in the model of ($m,s$)-particle. The complete integrability
of the dynamical equations with arbitrary Lagrange multipliers, in
spite of their nonlinearity, seems to be natural because of the fact
that the theory describes a free particle. To analyze the equations of
motion, it is helpful to introduce another parametrization in the
phase space.

Let us consider the domain ${\tilde T}^*(S^2)$ in the cotangent bundle
of sphere $T^*(S^2)$, which is selected out by the condition $p_z \neq
0$. In accordance with the restriction (31.b), the spherical part of
the constrained surface in the phase space is embedded completely in
${\tilde T}^*(S^2)$. On ${\tilde T}^*(S^2)$ one can replace the phase
variables $z$ and $p_z$ by a covariant parametrization in terms of the
coordinates
$$
q^\alpha = z^\alpha\sqrt{2p_z}
\eqno{(39)}$$
transforming by the spinor law (A.3) under the Lorentz group. As is
easily seen, the correspondence $(z,p_z) \rightarrow q^\alpha$ defined
by Eq. (39) maps the space ${\tilde T}^*(S^2)$ onto ${\Bbb C}^2_*$ (see
Appendix A); then ${\tilde T}^*(S^2)$ turns out to be the factor-space
of ${\Bbb C}^2_*$ with respect to the equivalence relation $q^\alpha
\sim -q^\alpha$. The local functions $q^\alpha$ (39) on the phase space
and their conjugates ${\bar q}^{\dot\alpha}$ possess the following
Poisson brackets
$$
\{ q^\alpha ,q^\beta \} = -\varepsilon^{\alpha\beta}, \qquad \{
{\bar q}^{\dot\alpha},{\bar q}^{\dot\beta} \} =
-\varepsilon^{{\dot\alpha}{\dot\beta}}, \qquad \{ q^\alpha , {\bar
q}^{\dot\beta} \} = 0,
\eqno{(40)}$$
$\varepsilon^{\alpha\beta}$ and $\varepsilon^{{\dot\alpha}{\dot\beta}}$
being the spinor metrics $(\varepsilon^{01} = \varepsilon^{{\dot
0}{\dot 1}} = 1)$. These relations imply that the transformation (39)
is canonical.

In terms of the variables $q^\alpha$ and ${\bar q}^{\dot\alpha}$, the
constraint (31.b) looks like
$$
T_2 = \frac{1}{4}(p^a{\cal F}_a)^2 - \Delta^2, \qquad {\cal F}_a =
(\sigma_a)_{\alpha{\dot\alpha}}q^\alpha {\bar q}^{\dot\alpha},
\eqno{(41)}$$
and the Hamilton actions (34) takes the explicitly Lorentz-covariant
form
$$
S = \int {\rm d}\tau (p_a{\dot x}^a + \frac{1}{2} \{ q_\alpha{\dot
q}^\alpha + {\bar q}_{\dot\alpha} {\dot{\bar q}}^{\dot\alpha} -
\lambda_iT_i\}).
\eqno{(42)}$$
Let us introduce the 4-component Majorana spinor
$$
Q = \left( \begin{array}{l} q_\alpha\\
{\bar q}^{\dot\alpha} \end{array} \right), \qquad {\overline Q} =
(q^\alpha , {\bar q}_{\dot\alpha} )
\eqno{(43)}$$
and make use of the identities
$$
{\cal F}^a = \frac{1}{2} {\overline Q}\gamma^aQ, \qquad q^\alpha {\dot
q}_\alpha + {\bar q}^{\dot\alpha} {\dot{\bar q}}_{\dot\alpha} =
{\overline Q}\gamma_5{\dot Q}.
$$
Then, the Hamilton equations can be written as follows:
$$
{\dot p}^a = 0, \qquad {\dot x}^a - \lambda_1p^a - \frac{1}{4}
\lambda_2(p,{\cal F}){\cal F}^a = \vspace{-12pt}0,
$$
{~} \hfill \vspace{-12pt}(44)\\
$$
{\dot Q} + \frac{1}{4}\lambda_2(p,{\cal F})\gamma_5(p,\gamma )Q = 0.
$$
Taking into account the constraints, these equations are easily
integrated for arbitrary $\lambda_1(\tau)$ and $\lambda_2(\tau)$. The
general solution reads
$$
Q(\tau ,\tau_0) = \exp \Big[-\alpha (\tau ,\tau_0)\gamma_5(p,\gamma
\vspace{-12pt})\Big]Q(\tau_0),
$$
{~}\hfill \vspace{-12pt}(45.a)\\
$$
\alpha (\tau ,\tau_0) = \frac{\Delta}{2} \int_{\tau_0}^\tau {\rm d}s\,
\lambda_2(s);
$$
$$
x^a(\tau ,\tau_0) = x^a(\tau_0) + p^a \int_{\tau_0}^\tau {\rm d}s
\left\{ \lambda_1(s) + \left( \frac{\Delta}{m} \right)^2 \lambda_2(s)
\Big[\cos \Big(2m\alpha (s,\tau_0) \Big)- 1\Big]\right\} 
$$
$$
+ \frac{\Delta}{2}{\cal F}^a(\tau_0) \int_{\tau_0}^\tau {\rm d}s~
\lambda_2(s)\, \cos \Big(2m\alpha (s,\tau_0)\Big)
$$
$$
+\frac{\Delta}{8m}  \overline{Q} (\tau_0)\gamma_5 \big[ \gamma^a, \gamma^b \big] 
Q(\tau_0)\, p_b \int_{\tau_0}^\tau {\rm d}s\,
\lambda_2(s)\, \sin \Big(2m\alpha (s,\tau_0) \Big).
\eqno{(45.b)}$$
To restore the general solution in the gauge (20), one is simply to
use the correspondence $e_1 \leftrightarrow \lambda_1$, $e_2
\leftrightarrow \lambda_2$ together with the relation $z(\tau ,\tau_0)
= q^1(\tau ,\tau_0)/q^0(\tau ,\tau_0)$.

\section{Quantization of the ($m,s$)-particle model}

In the present section, a realization of operator formulation is
suggested for the quantum theory of ($m,s$)-particle. The phase
variables in this formulation are considered to be Hermitian operators
defined in a Hilbert space, with Poincar\'e-invariant inner product,
and subjected to the canonical commutation relations. Operator ordering
in functions on the phase space is chosen in such a way that associated
with the Hamilton generators (36) of the Poincar\'e group will be
Hermitian operators ${\Bbb P}_a$ and ${\Bbb J}_{ab}$ satisfying the
commutation relations
$$
[{\Bbb P}_a, {\Bbb P}_b] = 0, \qquad [{\Bbb J}_{ab}, {\Bbb P}_c] = {\rm
i}\eta_{ac}{\Bbb P}_b - \vspace{-12pt}{\rm i}\eta_{bc}{\Bbb P}_a,
$$
{~} \hfill \vspace{-12pt}(46)\\
$$
[{\Bbb J}_{ab}, {\Bbb J}_{cd}] = {\rm i}\eta_{ac}{\Bbb J}_{bd} - {\rm
i}\eta_{bc}{\Bbb J}_{ad} + {\rm i}\eta_{ad}{\Bbb J}_{cb} - {\rm
i}\eta_{bd}{\Bbb J}_{ca}.
$$
Then, since the constraints (31) were expressed in terms of the
Hamilton generators, owing to Eq. (37), the quantum analogs of $T_i$
should be Hermitian operators of the form
$$
{\widehat T}_1 = {\Bbb P}^a{\Bbb P}_a + m^2{\mathbbm 1};
\eqno{(47.a)}$$
$$
{\widehat T}_2 = {\Bbb W}^a{\Bbb W}_a - m^2s(s+1){\mathbbm 1}, \qquad {\Bbb
W}^a = \frac{1}{2}\varepsilon^{abcd} {\Bbb P}_b{\Bbb J}_{cd}.
\eqno{(47.b)}$$
The physical states are to be subject to the conditions
$$
{\widehat T}_i|\Psi_{\rm phys}> = 0, \qquad i = 1,2,
\eqno{(47.c)}$$
which have the meaning that the Casimir operators of the Poincar\'e
group are multiples of unity on the space of physical states.

As we have seen, the general structure of physical observables in
the model of ($m,s$)-particle is described by Eq. (38). It is now
obvious that the quantization procedure formulated allows to assign to
every classical physical observable a well defined Hermitian operator
in the Hilbert space.

In the classical regime, the parameter $s$ could take arbitrary
non-negative values, and for any such value the dynamics was
non-contradictory. It turns out, however, that nontrivial solutions to
the equations for physical states exist only for (half)integer $s$.
Really, the set of Eqs. (46), (47) coincides with that specifying an
irreducible massive spin-$s$ unitary representation of the Poincar\'e
group. As is well known, such representations exist only for
(half)integer spins.

The problem of constructing the operator formulation described reduces
to finding an operator realization for the canonical commutation
relations consistent with the conditions (46), (47). So, the problem
is in finding realizations for the massive irreducible Poincar\'e
representations in spaces of tensor fields on ${\cal M}^6$.

The general structure of tensor fields on ${\cal M}^6$ is discussed in
Appendix B. Here we will be interested only in those tensor fields on
${\cal M}^6$ which are scalar in Minkowski space and have a spherical
type $\{r/2,s/2\}$, $r$ and $s$ being integers. The
Poincar\'e-generators for the fields of type $\{r/2,s/2\}$ are given by
Eqs. (B.8) and (B.9).

Let us consider the space $^\uparrow{\cal H}^{[n/2]}({\cal M}^6;m)$ of
massive positive-frequency fields on ${\cal M}^6$ of type $[n/2] \equiv
\{n/2,0\}$, where $n = 0, 1, 2, \ldots$. Such fields are subject to
the mass-shell condition (47.a),
$$
(\Box - m^2)\Phi^{[n/2]}(x,z,{\bar z}) = 0,
\eqno{(48)}$$
and possess the Fourier decomposition
$$
\Phi^{[n/2]}(x,z,{\bar z}) = \int\frac{{\rm d}^3{\vec p}}{p^0} {\rm
e}^{{\rm i}(p,x)}\Phi^{[n/2]}(p,z,{\bar z}\vspace{-12pt}),
$$
{~}\hfill \vspace{-12pt}(49)\\
$$
p^2 + m^2 = 0, \qquad p^0 > 0.
$$
We define inner product for the elements from $^\uparrow{\cal
H}^{[n/2]}({\cal M}^6;m)$ by the rule
$$
\ll\Phi_1|\Phi_2\gg_{[n/2]} = N \int\frac{{\rm d}^3{\vec p}}{p^0}~
\frac{{\rm d}z{\rm d}{\bar z}}{(p,\xi)^2}~\frac{1}{(p,\xi)^n} {\overline
{\Phi_1^{[n/2]}(p,z,{\bar z})}} \Phi_2^{[n/2]}(p,z,{\bar z}),
\eqno{(50)}$$
for $N$ being some normalization constant, $\xi_a$ given as in Eq. (6).
Then $^\uparrow{\cal H}^{[n/2]}({\cal M}^6;m)$ becomes a Hilbert space.
It should be stressed that the inner product (50) is well defined on
$S^2$ and built upon the metric (5). We note also that the integration
measure
$$
\frac{{\rm d}^3{\vec p}}{p^0}~\frac{{\rm d}z{\rm d}{\bar z}}{(p,\xi)^2}
$$
proves to be Poincar\'e invariant, and the expression
$$
(p,\xi)^{-n} {\overline{\Phi_1^{[n/2]}(p,z,{\bar z})}}
\Phi_2^{[n/2]}(p,z,{\bar z}),
$$
is a scalar with respect to the Poincar\'e group. As a result, the
Poincar\'e representation acting on $^\uparrow{\cal H}^{[n/2]}({\cal
M}^6;m)$ is unitary. This representation can be readily decomposed into
a direct sum of irreducible ones by accounting the fact that the spin
operator $C^{[n/2]} = {\Bbb W}^a{\Bbb W}_a$ on $^\uparrow{\cal
H}^{[n/2]}({\cal M}^6;m)$ coincides with the Laplacian $\Delta^{[n/2]}$
(B.11). The spectrum of $\Delta^{[n/2]}$ is described in Appendix C and
given by Eq. (C.1), where $R = -p^2 = m^2$ is the curvature of the
metric (5) building up from the 4--momentum. Thus we get the
decomposition
$$
^\uparrow{\cal H}^{[n/2]}({\cal M}^6;m) {\mathop{\oplus} \limits_
{s=\frac{n}{2},\frac{n}{2}+1,\ldots}}~ ^\uparrow{\cal
H}_s^{[n/2]}({\cal M}^6;m).
\eqno{(51)}$$
Here the invariant subspace $^\uparrow{\cal H}_s^{[n/2]}({\cal M}^6;m)$
realizes the Poincar\'e representation of mass $m$ and spin $s$, that
is, it is characterized by the conditions (47.a) and (47.b). As is shown
in Appendix C, the expansion of an arbitrary field from $^\uparrow{\cal
H}_s^{[n/2]}({\cal M}^6;m)$, which corresponds to the decomposition
(51), reads as follows
$$
\Phi^{[n/2]}(p,z,{\bar z}) = \sum_{s = \frac{n}{2},\frac{n}{2} + 1,
\ldots} F_{\alpha_1\ldots\alpha_{s+n/2}{\dot\alpha}_1\ldots
{\dot\alpha}_{s-n/2}}(p)\frac{z^{\alpha_1} \ldots z^{\alpha_{s+n/2}}
{\bar z}^{{\dot\alpha}_1} \ldots {\bar z}^{{\dot\alpha}_{s-n/2}}}
{(p,\xi)^{s-n/2}}.
\eqno{(52)}$$
In this expansion, each coefficient $F_{\alpha(s+n/2){\dot\alpha}(s-
n/2)}(p)$ is symmetric in its undotted and, independently, in its
dotted indices
$$
F_{\alpha_1\ldots\alpha_{s+n/2}{\dot\alpha}_1\ldots
{\dot\alpha}_{s-n/2}}(p) = F_{(\alpha_1\ldots\alpha_{s+n/2})
({\dot\alpha}_1\ldots {\dot\alpha}_{s-n/2})}(p),
\eqno{(53)}$$
and $p$-transversal,
$$
p^{\beta\dot\beta} F_{\beta\alpha(s+n/2-1){\dot\beta}{\dot\alpha}(s-
n/2-1)}(p) = 0.
\eqno{(54)}$$
The $F$'s can be identified with the Fourier transforms of tensor
fields on Minkowski space.

We describe in more detail the case of massive scalar fields on
${\cal M}^6$. As is seen from (51), the corresponding representation of
the Poincar\'e group on  $^\uparrow{\cal H}^{[0]}({\cal M}^6;m)$ is
decomposed into the direct sum of all representations with integer
spins. Therefore, a massive scalar field on ${\cal M}^6$ generates
massive fields of arbitrary integer spins in Minkowski space. For $n$ =
0 the decomposition (52) can be rewritten on the manner
$$
\Phi^{[0]}(p,z,{\bar z}) = \sum_{s=0}^\infty F_{a_1\ldots a_s}(p)
\frac{\xi^{a_1}\ldots\xi^{a_s}}{(p,\xi)^s},
\eqno{(55)}$$
where
$$
F_{a_1\ldots a_s}(p) = \left(-\frac{1}{2}\right)^s ({\tilde\sigma}_
{a_1})^{{\dot\alpha}_1\alpha_1} \ldots ({\tilde\sigma}_{a_s})^
{{\dot\alpha}_s\alpha_s} F_{\alpha_1\ldots\alpha_s{\dot\alpha}_1
\ldots{\dot\alpha}_s}(p).
$$
For just introduced fields with vector indices, the requirement (53)
means that $F_{a(s)}$ is symmetric and traceless
$$
F_{a_1\ldots a_s}(p) = F_{(a_1\ldots \vspace{-12pt}a_s)}(p),
$$
{~}\hfill \vspace{-12pt}(56)\\
$$
{F^b}_{ba_1\ldots a_{s-2}}(p) = 0,
$$
while the condition (54) takes the form
$$
p^bF_{ba_1\ldots a_{s-1}}(p) = 0.
\eqno{(57)}$$
Together with the mass-shell condition $p^2 + m^2 = 0$, the equations
(56) and (57) constitute the set of relativistic wave equations for a
massive field of integer spin $s$ $^{19}$.

It seems instructive to reexpress the scalar product (51) for $n = 0$
in terms of the fields appearing in the expansion (55). Direct
calculations basing on the use of the identity
$$
\int \frac{{\rm d}z{\rm d}{\bar z}}{(p,\xi)^2} = -~\frac{\pi}{p^2}
$$
lead to the following result
$$
\ll\Phi_1|\Phi_2\gg_{[0]} = \pi N \int\frac{{\rm d}^3{\vec p}}{p^0}
\sum_{s=0}^\infty \left(\frac{2}{m^2}\right)^s \frac{(s!)^2}{(2s+1)!}
{\overline{\Phi_1^{a_1\ldots a_s}(p)}} \Phi_{2~a_1\ldots a_s}(p).
\eqno{(58)}$$

Now, let us consider massive spinor fields on ${\cal M}^6$, i.e. $n =
1$. The relations (51) show that the Poincar\'e representation defined
on $^\uparrow{\cal H}^{[n/2]}({\cal M}^6;m)$ is decomposed into the
direct sum of all representations with half-integer spins. Thus, a
massive spinor field on ${\cal M}^6$ generates massive fields with
arbitrary half-integer spins. The decomposition (52) can be rewritten
for $n = 1$ by the rule
$$
\Phi^{[1/2]}(p,z,{\bar z}) = \sum_{k=0}^\infty
F_{a_1\ldots a_k\alpha}(p)\frac{z^\alpha\xi^{a_1} \ldots \xi^{a_k}}
{(p,\xi)^k},
\eqno{(59)}$$
where
$$
F_{a_1\ldots a_k\alpha}(p) = \left(-~\frac{1}{2}\right)^k
({\tilde\sigma}_{a_1})^{{\dot\beta}_1\beta_1}\ldots
({\tilde\sigma}_{a_k})^{{\dot\beta}_k\beta_k}
F_{\alpha\beta_1\ldots\beta_k{\dot\beta}_1 \ldots{\dot\beta}_k}.
$$
In terms of the spin-tensors introduced, the requirements (53) and (54)
are equivalent to the equations
$$
F_{a_1\ldots a_k\alpha}(p) = F_{(a_1\ldots a_k)\alpha}(p), \qquad
{F^b}_{ba_1\ldots a_{k-2}\alpha}(p) = \vspace{-12pt}0,
$$
{~}\hfill \vspace{-12pt}(60)\\
$$
({\tilde\sigma}^b)^{{\dot\beta}\beta} F_{ba_1\ldots a_{k-1}\beta}(p) =
p^bF_{ba_1\ldots a_{k-1}\alpha}(p) = 0,
$$
which form, together with the condition $p^2 + m^2 = 0$, the set of
relativistic wave equations determining an on-shell massive field of
half-integer spin $(k+1/2)$ $^{20}$. It is worth remarking that a
massive field $\Phi^{[1/2]}$ on ${\cal M}^6$ turns out to describe spin
1/2 if and only if $\Phi^{[1/2]}$ is holomorphic, $\partial_{\bar z}
\Phi^{[1/2]} = 0$.

Real massive fields on ${\cal M}^6$ can be realized by considering
spaces
$$
{\cal H}^{[n/2]}({\cal M}^6;m) =~^\uparrow{\cal H}^{[n/2]}({\cal
M}^6;m) \oplus~ ^\downarrow{\cal H}^{[n/2]}({\cal M}^6;m)
$$
spanned on fields with mixed frequency and, then, selecting out in
these spaces real Poincar\'e-invariant subspaces. For instance, a real
field $\Phi^{[1/2]}$ of spin 1/2 satisfies the equations
$$
(p,\xi)\nabla_z\Phi^{[1/2]} = m{\overline{\Phi^{[1/2]}}}, \qquad
\partial_{\bar z} \Phi^{[1/2]} = 0
$$
in momentum space, or
$$
\xi^a\stackrel{\leftrightarrow}{\partial_z}\partial_a\Phi^{[1/2]} =
{\rm i}m{\overline{\Phi^{[1/2]}}}, \qquad \partial_{\bar z}
\Phi^{[1/2]} = 0
$$
in coordinate space. The above equations prove to be equivalent to
ordinary Dirac equation on Majorana spinor field.

\section{Conclusion}

In the present paper we have constructed the model of relativistic
massive particle of arbitrary spin with the configuration space ${\cal
M}^6 = {\Bbb R}^{3,1}\times S^2$. The theory possesses two gauge
symmetries of reparametrization type. This gauge structure induces
strong conservation of the phase-space functions of squared momentum
and squared Pauli--Lubanski vector, while in the quantum theory it
implies that the Casimir operators of the Poincar\'e group are
multiples of unity in the space of physical states. That is why we can
identify one of the two parameters arising in the Lagrangian with the
spin of particle.

The theory suggested has quite clear and simple geometric origin, and
can be also considered as a minimal model of relativistic massive
spinning particle. The point is that dimension of the corresponding
configuration space turns out to be minimally possible to describe the
joint space-time evolution and dynamics of spin.

Our model admits a number of nontrivial generalizations. In particular,
we have already developed (super)particle models extending the model of
($m,s$)-particle to the cases when the space-time manifold ${\Bbb
R}^{3,1}$ is replaced by $N$-extended flat global superspace, (anti) de
Sitter space, anti--de Sitter superspace$^{20,21}$.

The model of this paper describes the dynamics of free spinning particle with
non-vanishing mass in terms of ${\cal M}^6$-geometry. It is believed,
however, that in the massless case total configuration space should
also involve sphere as a subspace. The appearance of sphere $S^{D-2}$
for describing massless dynamics in special dimensions $D$ = 3, 4, 6,
and 10 has recently been demonstrated for the case of a massless
superparticle by Galperin, Howe, and Stelle$^{22}$ suggested an
elegant group-theoretic interpretation for the twistor formulation of
superparticle$^{23}$. In the approach of Refs. 23, 24, there appeared a
light-like $D$-vector as a variable representing spherical variables.
This is analogous to the formulation (26) for our model. Interestingly,
the Lagrangian (26) leads to a consistent spinning particle model in
any dimension in the sense that the local transformations (27)
and (28) leave invariant the corresponding $D$-dimensional
action functional. But only for $D$ = 3, 4 the theory describes
a massive particle with arbitrary irreducible spin.

In conclusion, we would like to note an interesting relationship
between a special algebra of functions on the phase space of
($m,s$)-particle and higher spin superalgebras$^{24}$ underlying the
theories of interacting higher spin massless fields$^{25}$. As has been
shown in section 3, the spherical part of the constraint surface in the
phase space can be parametrized by the spinor variables (39). The set
of all regular functions of $q$ and ${\bar q}$,
$$
f(q,{\bar q}) = \sum_{s=0}^\infty \sum_{\begin{array}{c}\scriptstyle{
m,n \geq 0}\\
\scriptstyle{m+n=s} \end{array}} f_{\alpha_1\ldots\alpha_m{\dot\alpha}_1
\ldots {\dot\alpha}_n} q^{\alpha_1}\ldots q^{\alpha_m} {\bar
q}^{{\dot\alpha}_1}\ldots {\bar q}^{{\dot\alpha}_n},
$$
forms an infinite dimensional Lie algebra, with respect to the Poisson
bracket (39). The quantization of this algebra, which consists in
associating with the phase variables $q^\alpha$ and ${\bar
q}^{\dot\alpha}$ operators ${\hat q}^\alpha$ and ${\hat{\bar
q}}^{\dot\alpha}$ under the commutation relations
$$
[{\hat q}^\alpha,{\hat q}^\beta] = {\rm i}\varepsilon^{\alpha\beta},
\qquad [{\hat{\bar q}}^{\dot\alpha}, {\hat{\bar q}}^{\dot\beta}] = {\rm
i}\varepsilon^{{\dot\alpha}{\dot\beta}}, \qquad [{\hat q}^\alpha,
{\hat{\bar q}}^{\dot\alpha}] = 0
$$
leads to an associative operator algebra being naturally ${\Bbb
Z}_2$-graded and, hence, possessing the structure of superalgebra. The
superalgebra obtained proves to coincide with some special higher spin
superalgebra$^{25}$.

\section{Acknowledgements}

The authors are grateful to V.G. Bagrov, I.A. Batalin and A.A. Sharapov
for useful discussions. This work was supported in part by the
International Science Foundation under the Emergency Grants Program.

\bigskip

\centerline{{\bf Appendix A}. The Lorentz group action on $S^2$}

\bigskip

A two sphere $S^2 = {\Bbb C} \cup \{\infty\}$ is naturally endowed by
the structure of a transformation space for the Lorentz group
$SL(2,{\Bbb C})/{\Bbb Z}_2$ if one realizes $S^2$ as the complex
projective space ${\Bbb C}P^1$, i.e. the factor-space of the complex
space ${\Bbb C}^2_* = {\Bbb C}^2 \backslash \{(0,0)\}$, spanned by
non-zero complex two-vectors $q^\alpha = (q^0,q^1)$, with respect to
the equivalence relation $q^\alpha \sim \lambda q^\alpha, \forall~
\lambda \in {\Bbb C}_* = {\Bbb C} \backslash \{0\}$. Let $\pi: {\Bbb
C}^2_* \rightarrow S^2$ be the canonical projection. The open cover
of ${\Bbb C}^2_*$ by two charts
$$
U_0 = \{q^\alpha \in {\Bbb C}^2_*, q^0 \neq 0\}, \qquad z \equiv
\vspace{-12pt}q^1/q^0;
$$
{~}\hfill \vspace{-12pt}(A.1)\\
$$
U_1 = \{q^\alpha \in {\Bbb C}^2_*, q^1 \neq 0\}, \qquad w \equiv
-q^0/q^1
$$
induces the atlas $\{\pi(U_0),\pi(U_1)\}$ on $S^2$; the only point of
$S^2$ not covered by the chart $\pi(U_0)$ can be identified with
infinitely removed point. On the role of local complex coordinates in
the charts $\pi(U_0)$ and $\pi(U_1)$ it is useful to choose the
variables $z$ and $w$, respectively, connected to each other in the
overlap $\pi(U_0) \cap \pi(U_1) = {\Bbb C}_*$ by the transition
function $w = -1/z$.

Let us consider the spinor representation (1/2, 0) of the Lorentz
group, i.e. the representation of $SL(2,{\Bbb C})$ on ${\Bbb C}^2$
assigning to each matrix
$$
N = ({N_\alpha}^\beta) = \left( \begin{array}{cc} a & b\\
c & d \end{array} \right) \in SL(2,{\Bbb C}), \qquad \alpha,\beta = 0,1
\eqno{({\rm A}.2)}$$
the following transformation on ${\Bbb C}^2$
$$
q^\alpha \rightarrow {q'}^\alpha = q^\beta {(N^{-1})_\beta}^\alpha.
\eqno{({\rm A}.3)}$$
Since mutually equivalent points from ${\Bbb C}^2_*$ are moved into
equivalent ones, the restriction of this representation to ${\Bbb
C}^2_*$ indices some action of the group on $S^2$ given in the local
coordinates $z, {\bar z}$ by fractional linear transformations of the
form
$$
z \rightarrow z' = \frac{az-b}{-cz+d}.
\eqno{({\rm A}.4)}$$
For any $N \in SL(2,{\Bbb C})$, the group elements $\pm N$ define
the same transformation on $S^2$, therefore we result in the action of
the Lorentz group on $S^2$.

In accordance with Eq. (A.4), the Lorentz group acts on $S^2$ by
holomorphic fractional linear transformations. Reversally, it is easy
to observe that every holomorphic one--to--one mapping of $S^2$ onto
itself is a transformation of the form (A.4). Therefore, the Lorentz
group coincides with the group of complex automorphisms of
two-dimensional sphere.

\bigskip

\centerline{{\bf Appendix B}. Tensor fields on ${\cal M}^6$}

\bigskip

Representations of the Poincar\'e group in terms of fields living on
${\cal M}^6$ can be obtained via tensor products of analogous
representations defined on Minkowski space and on two-sphere. Let us
recall that a tensor field on ${\Bbb R}^{3,1}$ of Lorentz type $(k/2,
l/2),~k,l = 0,1,2,\ldots,$ is given by a set of smooth components
$$
F_{\alpha(k){\dot\alpha}(l)}(x) = F_{\alpha_1\ldots\alpha_k
{\dot\alpha}_1\ldots{\dot\alpha}_l}(x),
\eqno{({\rm B}.1)}$$
symmetric independently in their undotted and dotted indices,
$$
F_{\alpha_1\ldots\alpha_k{\dot\alpha}_1\ldots{\dot\alpha}_l}(x) =
F_{(\alpha_1\ldots\alpha_k)({\dot\alpha}_1\ldots{\dot\alpha}_l)}(x)
\eqno{({\rm B}.2)}$$
and changing by the law
$$
{F'}_{\alpha_1\ldots\alpha_k{\dot\alpha}_1\ldots{\dot\alpha}_l}(x') =
{N_{\alpha_1}}^{\beta_1} \ldots {N_{\alpha_k}}^{\beta_k}
\left.{\bar N}_{{\dot\alpha}_1}\right.^{{\dot\beta}_1} \ldots
\left.{\bar N}_{{\dot\alpha}_l}\right.^{{\dot\beta}_l}
F_{\beta_1\ldots\beta_k{\dot\beta}_1\ldots{\dot\beta}_l}(x)
\eqno{({\rm B}.3)}$$
under the Poincar\'e transformations (1). The round brackets in (B.2)
mean the symmetrization of indices. A tensor field on $S^2$ of type
$\{r/2,s/2\}$, where $r,s = 0, \pm 1, \pm 2, \ldots,$ is given by
a smooth function $\Phi(z,{\bar z})$ in chart ${\Bbb C}$ and by
a smooth function $\Psi(w,{\bar w})$ in chart ${\Bbb C}_* \cap
\{\infty\}$ such that in the overlap of charts these functions are
connected as follows:
$$
\Psi(w,{\bar w}) = \left(\frac{\partial w}{\partial z}\right)^{r/2}
\left(\frac{\partial{\bar w}}{\partial{\bar z}}\right)^{s/2} \Phi(z,{\bar
z}).
\eqno{({\rm B}.4)}$$
The Lorentz transformations (A.4) act on the field of type $\{r/2,
s/2\}$ by the law
$$
\Phi'(z',{\bar z}') = \left(\frac{\partial z'}{\partial z}\right)^{r/2}
\left(\frac{\partial{\bar z}'}{\partial{\bar z}}\right)^{s/2}
\Phi(z,{\bar z}).
\eqno{({\rm B}.5)}$$
In summary, each tensor field on ${\cal M}^6$ is characterized by its
Lorentz type $\{k/2, l/2\}$ and its spherical type $\{r/2, s/2\}$.

In the present paper, we mainly consider tensor fields on ${\cal M}^6$
without external Lorentz indices $(k = l= 0)$. For most of applications
this proves to be sufficient, because ordinary tensor fields in
Minkowski space arise as coefficients in expansions of tensor fields on
${\cal M}^6$ over special harmonics. Let us consider, for instance, a
tensor field on $S^2$ of type $\{n/2, 0\} \equiv [n/2]$, $\Phi^{[n/2]}$,
satisfying the requirement of holomorphicity
$$
\partial_{\bar z}\Phi^{[n/2]} = 0.
\eqno{({\rm B}.6)}$$
According to the Riemann---Roch theorem$^{18}$, Eq. (B.6) possesses
nontrivial solutions on $S^2$ only for $n \geq 0$ (in the class of
smooth tensor fields on $S^2$), and dimension of the corresponding
space of holomorphic fields is equal to $(n+1)$. The general solution
of Eq. (B.6) reads
$$
\Phi^{[n/2]}(z) = F_{\alpha_1\ldots\alpha_n}z^{\alpha_1}\ldots
\vspace{-12pt}z^{\alpha_n},
$$
{~}\hfill \vspace{-12pt}(B.7)\\
$$
\Rightarrow \Psi^{[n/2]}(w) = F_{\alpha_1\ldots\alpha_n}w^{\alpha_1}
\ldots w^{\alpha_n}, \qquad w^\alpha = (w, -1).
$$
Here $F_{\alpha_1\ldots\alpha_n}$ is a Lorentz tensor of type $(n/2,
0)$.

Representation of the Poincar\'e group in the space of tensor fields on
${\cal M}^6$ of type $\{r/2, s/2\}$ is characterized by the following
generators:
$$
{\Bbb P}_a = -{\rm i} \partial_a, \qquad
{\Bbb J}_{ab} = -{\rm i}(x_a\partial_b - x_b\partial_a + M_{ab}),
\eqno{({\rm B}.8)}$$
where the spinning part of ${\Bbb J}_{ab}$ is realized by spherical
variables on the manner
$$
M_{ab} = (\sigma_{ab})_{\alpha\beta}M^{\alpha\beta} -
({\tilde\sigma}_{ab})_{{\dot\alpha}{\dot\beta}} \vspace{-12pt}{\bar
M}^{{\dot\alpha}{\dot\beta}},
$$
$$
\vspace{-12pt}{~}
\eqno{({\rm B}.9)}$$
$$
M^{\alpha\beta} = -z^\alpha z^\beta\partial_z + \frac{r}{2} \partial_z
(z^\alpha z^\beta), \qquad {\bar M}^{{\dot\alpha}{\dot\beta}} = -
{\bar z}^{\dot\alpha} {\bar z}^{\dot\beta}\partial_{\bar z} +
\frac{s}{2} \partial_{\bar z}({\bar z}^{\dot\alpha} {\bar
z}^{\dot\beta}).
$$
Then, the operator of squared spin $C^{\{r/2, s/2\}} = {\Bbb W}^a {\Bbb
W}_a$, ${\Bbb W}^a = \frac{1}{2}\varepsilon^{abcd}{\Bbb P}_b {\Bbb
J}_{cd}$ being the Pauli---Lubanski vector, has the form
$$
C^{\{r/2, s/2\}} = -({\Bbb P},\xi)^2\partial_z\partial_{\bar z} +
r({\Bbb P},\xi)({\Bbb P},\partial_z\xi)\partial_{\bar z} + s({\Bbb
P},\xi)({\Bbb P},\partial_{\bar z}\xi)\partial_z -
$$
$$
- rs({\Bbb P}, \partial_z \xi)({\Bbb P},\partial_{\bar z}\xi) - {\Bbb P}^2
\left\{ \left(\frac{r-s}{2}\right)^2 + \frac{r+s}{2}\right\}.
\eqno{({\rm B}.10)}$$
Let us restrict this representation to the subspace of massive fields
satisfying the Klein---Gordon equation ${\Bbb P}^2 + m^2{\mathbbm 1} = 0$
and introduce the Fourier transform. In result, the spin operator
(B.10) can be expressed in terms of the covariant derivatives
constructed on the base of the metric (5). In particular, for the
massive tensor fields on ${\cal M}^6$ of type $\{n/2, 0\} \equiv [n/2]$
we get
$$
C^{[n/2]} \equiv C^{\{n/2,0\}} = \vspace{-12pt}\Delta^{[n/2]},
$$
$$
\vspace{-12pt}{~}
\eqno{({\rm B}.11)}$$
$$
\Delta^{[n/2]} = -2g^{z{\bar z}}\nabla_z\nabla_{\bar z} + \frac{n}{2}
\left(\frac{n}{2} + 1\right)R.
$$
As is seen, the operator of squared spin is determined by special
spherical Laplacian.

\bigskip

\centerline{{\bf Appendix C}. Relativistic harmonic analysis on $S^2$}

\bigskip

In this appendix we shall prove a lot of important assertions
concerning spectra of the Laplacians (B.11) and generalized Fourier
decompositions for tensor fields on $S^2$. We begin with formulating
the basic statements.

I. Spectrum of the Laplacian $\Delta^{[n/2]}$, where $n = 0,1,2,
\ldots,$ is given by the following eigenvalues
$$
s(s+1)R, \qquad s = \frac{n}{2},\frac{n}{2}+1, \frac{n}{2}+2, \ldots;
\eqno{({\rm C}.1)}$$
dimension of the eigenspace corresponding to an eigenvalue $s(s+1)R$ is
equal to $(2s+1)$.

II. A smooth tensor field $\Phi^{[n/2]}(z,{\bar z})$ on $S^2$ of type
$[n/2] \equiv \{n/2, 0\}, n \geq 0$, can be represented in the form
$$
\Phi^{[n/2]}(z,{\bar z}) = \sum_{k=0}^\infty F_{\alpha_1\ldots
\alpha_{n+k}{\dot\alpha}_1\ldots{\dot\alpha}_k} \frac{z^{\alpha_1}
\ldots z^{\alpha_{n+k}} {\bar z}^{{\dot\alpha}_1} \ldots {\bar
z}^{{\dot\alpha}_k}} {(p,\xi)^k}.
\eqno{({\rm C}.2)}$$
Here the expansion coefficients, being Lorentz tensors, are determined
uniquely from the two requirements:\\
a) $F_{\alpha_1\ldots \alpha_{n+k}{\dot\alpha}_1\ldots{\dot\alpha}_k}$
is a tensor of type $((n+k)/2, k/2)$, i.e.
$$
F_{\alpha_1\ldots \alpha_{n+k}{\dot\alpha}_1\ldots{\dot\alpha}_k} =
F_{(\alpha_1\ldots \alpha_{n+k})({\dot\alpha}_1\ldots{\dot\alpha}_k});
\eqno{({\rm C}.3)}$$
b) for $k \neq 0$, $F_{\alpha_{n+k}{\dot\alpha(k)}}$ is $p$-transversal,
$$
p^{\beta\dot\beta} F_{\beta\alpha(n+k-1){\dot\beta}{\dot\alpha}(k-1)}
= 0.
\eqno{({\rm C}.4)}$$
III. Under the fulfilment of Eqs. (C.3) and (C.4), the $k$-th term in
the expansion (C.2) is an eigenvector for $\Delta^{[n/2]}$ with the
eigenvalue $s(s+1)R$, where $s = n/2 + k$.

To prove the assertions formulated, we associate with each non-negative
integer $n$ the Hilbert space ${\cal H}^{[n/2]}$ of squared integrable
tensor fields on $S^2$ of type $[n/2]$ with respect to the inner
product
$$
<\Phi_1^{[n/2]}|\Phi_2^{[n/2]}> \equiv <\Phi_1|\Phi_2>_{[n/2]} = \int
{\rm d}z{\rm d}{\bar z} (g_{z{\bar z}})^{n/2+1}{\overline
{\Phi_1(z,{\bar z})}} \Phi_2(z,{\bar z}),
\eqno{({\rm C}.5)}$$
$\Phi_1$ and $\Phi_2$ being tensor fields on $S^2$ of type $[n/2]$.
One immediately gets the following identities:
$$
<\nabla_z\Phi^{[n/2+1]}|\Phi^{[n/2]}> = - <\Phi^{[n/2+1]}|\nabla^z
\Phi^{[n/2]}>;
\eqno{({\rm C}.6.a)}$$
$$
<\nabla^z\Phi^{[n/2]}|\Phi^{[n/2+1]}> = - <\Phi^{[n/2]}|\nabla_z
\Phi^{[n/2+1]}>;
\eqno{({\rm C}.6.b)}$$
$$
<\Phi|\Delta^{[n/2]}|\Phi>_{[n/2]} = 2\|\nabla^z\Phi\|^2_{[n/2+1]} +
\frac{n}{2}\left(\frac{n}{2}+1\right) \|\Phi\|^2_{[n/2]},
\eqno{({\rm C}.6.c)}$$
where $\nabla^z \equiv g^{z{\bar z}}\nabla_{\bar z}$. The final
identity shows that $\Delta^{[n/2]}$ is a positive operator on ${\cal
H}^{[n/2]}$,
$$
<\Phi|\Delta^{[n/2]}|\Phi>_{[n/2]} \geq \frac{n}{2}\left(\frac{n}{2}+
1\right) \|\Phi\|^2_{[n/2]},
$$
and the equality here proves to take place only on the subspace of
holomorphic fields
$$
{\cal H}_0^{[n/2]} = \left\{\Phi^{[n/2]} \in {\cal H}^{[n/2]}, \quad
\nabla^z\Phi^{[n/2]} = g^{z{\bar z}} \partial_{\bar z}\Phi^{[n/2]} = 0
\right\}.
\eqno{({\rm C}.7)}$$
Explicit structure of the states from ${\cal H}_0^{[n/2]}$ is given by
Eq. (B.7), and each such field is seen to be an eigenstate for
$\Delta^{[n/2]}$ with the eigenvalue $\frac{n}{2}\left(\frac{n}{2}+
1\right)R$. Obviously, this is the eigenvalue $\frac{n}{2}\left(
\frac{n}{2}+ 1\right)R$ realizing minimum in the spectrum of
$\Delta^{[n/2]}$. The rest eigenvalues of $\Delta^{[n/2]}$ can be
readily restored by taking into account two simple identities
$$
\nabla^z\Delta^{[n/2]} = \Delta^{[n/2+1]} \vspace{-12pt}\nabla^z,
$$
{~}\hfill \vspace{-12pt}(C.8)\\
$$
\nabla_z \Delta^{[n/2+1]} = \Delta^{[n/2]}\nabla_z
$$
and analysing also the following mappings
$$
\nabla^z:~{\cal H}^{[n/2]} \rightarrow {\cal H}^{[n/2+1]}, \qquad
\nabla_z:~{\cal H}^{[n/2+1]} \rightarrow {\cal H}^{[n/2]}.
\eqno{({\rm C}.9)}$$
One can check that kernel (Ker) and total image (Im) for the mappings
(C.9) look like:
$$
{\rm Ker}~\nabla^z|_{{\cal H}^{[n/2]}} = {\cal H}_0^{[n/2]}, \qquad
{\rm Im}~\nabla^z|_{{\cal H}^{[n/2]}} = {\cal H}^{[n/2+1]};
\eqno{({\rm C}.10.a)}$$
$$
{\rm Ker}~\nabla_z|_{{\cal H}^{[n/2+1]}} = 0, \qquad {\rm Im}~
\nabla_z|_{{\cal H}^{[n/2+1]}} = {\cal H}^{[n/2]} \backslash {\cal
H}_0^{[n/2]}.
\eqno{({\rm C}.10.b)}$$
For example, let $\Phi_\perp^{[n/2+1]}$ be a field orthogonal to the
subspace $\nabla^z{\cal H}^{[n/2]}$ in ${\cal H}^{[n/2+1]}$. Then we
have
$$
0 =~ <\nabla^z \Phi^{[n/2]}|\Phi_\perp^{[n/2+1]}>~ =~ -<\Phi^{[n/2]}|
\nabla_z \Phi_\perp^{[n/2+1]}>,
$$
for any $\Phi^{[n/2]} \in {\cal H}^{[n/2]}$, hence $\nabla_z
\Phi_\perp^{[n/2+1]} = 0$, and therefore $\Phi_\perp^{[n/2+1]} = 0$.
Now, it is easy to deduce the assertion I from the relations (C.8) and
(C.10).

To check the assertions II and III, one is to account the explicit
structure of states from ${\cal H}^{[n/2]}$ and to calculate the action
of $\nabla_z$ and $\nabla^z$ on separate terms in the expansion (C.2).
Making use of the identities
$$
z^\alpha\partial_z z^\beta - z^\beta\partial_z z^\alpha =
\vspace{-12pt}\epsilon^{\alpha\beta},
$$
{~}\hfill \vspace{-12pt}(C.11)\\
$$
\nabla_z z^\alpha = -~\frac{{p^\alpha}_{\dot\beta}{\bar
z}^{\dot\beta}}{(p,\xi)}, \qquad \nabla_z {\bar z}^{\dot\alpha}
= \nabla_z \frac{1}{(p,\xi)} = 0,
$$
one gets
$$
{\nabla^z} F_{\alpha_1\ldots \alpha_{n+k}{\dot\alpha}_1\ldots
{\dot\alpha}_k} \frac{z^{\alpha_1} \ldots z^{\alpha_{n+k}} {\bar
z}^{{\dot\alpha}_1} \ldots {\bar z}^{{\dot\alpha}_k}} {(p,\xi)^k} =
$$
$$
= F_{\alpha_1\ldots \alpha_{n+k+1}{\dot\alpha}_1\ldots{\dot\alpha}_{k-1}}
\frac{z^{\alpha_1} \ldots z^{\alpha_{n+k+1}} {\bar z}^{{\dot\alpha}_1}
\ldots {\bar z}^{{\dot\alpha}_{k-1}}} {(p,\xi)^{k-1}}.
\eqno{({\rm C}.12.a)}$$
$$
F_{\alpha_1\ldots \alpha_{n+k+1}\dot\alpha_1\ldots\dot\alpha_{k-1}}
\equiv -~\frac{k}{2} {p_{\alpha_{n+k+1}}}^{\dot\beta} F_{\alpha_1\ldots
\alpha_{n+k}{\dot\alpha}_1\ldots{\dot\alpha}_k{\dot\beta}};
$$
$$
\nabla_z F_{\alpha_1\ldots \alpha_{n+k+1}{\dot\alpha}_1\ldots
{\dot\alpha}_{k-1}} \frac{z^{\alpha_1} \ldots z^{\alpha_{n+k+1}} {\bar
z}^{{\dot\alpha}_1} \ldots {\bar z}^{{\dot\alpha}_{k-1}}}
{(p,\xi)^{k-1}} =
$$
$$
F_{\alpha_1\ldots \alpha_{n+k}{\dot\alpha}_1\ldots
{\dot\alpha}_k} \frac{z^{\alpha_1} \ldots z^{\alpha_{n+k}} {\bar
z}^{{\dot\alpha}_1} \ldots {\bar z}^{{\dot\alpha}_k}} {(p,\xi)^k},
\eqno{({\rm C}.12.b)}$$
$$
F_{\alpha_1\ldots\alpha_{n+k}{\dot\alpha}_1\ldots {\dot\alpha}_k}
\equiv -(n+k+1){p^\beta}_{{\dot\alpha}_k}F_{\beta\alpha_1\ldots
\alpha_{n+k}{\dot\alpha}_1\ldots {\dot\alpha}_{k-1}}.
$$

As is clear from the above discussion, Eq. (C.2) constitutes the
decomposition of a tensor field $\Phi^{[n/2]}$ with respect to the
complete set of eigenfunctions associated with the elliptic operator
$\Delta^{[n/2]}$ (B.11). Since this Laplacian is specified by a
time-like 4-vector $p^a$, and the coefficients $F_{\alpha(n+k)
{\dot\alpha}(k)}$ in (C.2) are Lorentz tensors, the above decomposition
can be called an expansion over relativistic harmonics. Obviously, this
expansion appears to be most adapted for making explicitly Lorentz
covariant calculations. It is instructive to establish relationship
between relativistic and ordinary spherical harmonics. Let us consider,
for instance, a scalar field $\Phi(z,{\bar z})$ on $S^2$, for which Eq.
(C.2) reads
$$
\Phi(z,{\bar z)} = \sum_{s=0}^\infty \Phi_s(z,{\bar z}),
\eqno{({\rm C}.13)}$$
where the $s$-th term
$$
\Phi_s(z,{\bar z)} = F_{\alpha_1\ldots \alpha_s{\dot\alpha}_1\ldots
{\dot\alpha}_s} \frac{z^{\alpha_1} \ldots z^{\alpha_s} {\bar
z}^{{\dot\alpha}_1} \ldots {\bar z}^{{\dot\alpha}_s}} {(p,\xi)^s}
\eqno{({\rm C}.14)}$$
satisfies the equation
$$
\Delta^{[0]}\Phi_s = s(s+1)R\Phi_s,
\eqno{({\rm C}.15)}$$
what implies the fulfilment of Eqs. (C.3), (C.4) for the coefficients
$F_{\alpha(s){\dot\alpha}(s)}$. We choose a coordinate system where the
metric (5) is determined by 4--vector $p^a = (\sqrt{R}, 0, 0, 0)$ and,
hence, proportional to the standard metric on $S^2$,
$$
{\rm d}s^2 = R\frac{4{\rm d}z{\rm d}{\bar z}}{(1 + z{\bar z})^2}.
\eqno{({\rm C}.16)}$$
Let us also replace the variables $z,{\bar z}$ by standard spherical
angles $\theta,\varphi$. Their connection reads
$$
\cos~\theta = \frac{1 - z{\bar z}}{1 + z{\bar z}}, \qquad {\rm
e}^{2{\rm i}\varphi} = z/{\bar z},
\eqno{({\rm C}.17)}$$
and $\Delta^{[0]}$ takes the form
$$
\Delta^{[0]} = -R\Delta, \qquad \Delta = \frac{1}{\sin^2\theta}
\frac{\partial^2}{\partial\varphi^2} + \frac{1}{\sin~\theta}
\frac{\partial}{\partial\theta}\left(\sin~\theta
\frac{\partial}{\partial\theta}\right).
$$
As is well known, the general solution of the equation $\Delta\Phi_s =
-s(s+1)\Phi_s$ can be written as follows:
$$
\Phi_s(\theta,\varphi) = \sum_{m=-s}^s C_m {\rm e}^{{\rm i}m\varphi}
P_s^{|m|}(\cos~\theta),
\eqno{({\rm C}.18)}$$
with
$$
P_s^{|m|}(\cos~\theta) = \frac{1}{2^ss!} \sin^m\theta \frac{{\rm
d}^{s+m}}{({\rm d}\cos~\theta)^{s+m}}(\cos^2\theta - 1)^s
$$
being adjoint Legendre polinomials. Expressing $\theta,\varphi$ in
(C.18) via $z,{\bar z}$, with the use of (C.17), we arrive at
$$
\Phi_s(z,{\bar z}) = \Phi_s(\theta(z,{\bar z}), \varphi(z,{\bar z})) =
\frac{\pi^s(z,{\bar z})}{(1 + z{\bar z})^s},
$$
where $\pi^s$ is a polinomial of $z$ and ${\bar z}$ of general degree
$2s$. The final expression is in agreement with Eq. (C.14).

\newpage
\centerline{References}

\bigskip

\noindent
1. F.A. Berezin and M.S. Marinov, JETP Lett. {\bf 21}, 678 (1975);
Ann. Phys. {\bf 104}, 336 (1977).\\
2. A. Barducci, R. Casalbuoni, and L. Lusanna, Nuovo Cim. {\bf A35}, 377
(1976).\\
3. L. Brink, S. Deser, B. Zumino, P. Di Vecchia, and P.S. Howe, Phys.
Lett. {\bf B64}, 435 (1976).\\
4. L. Brink, P. Di Vecchia, and P.S. Howe, Nucl. Phys. {\bf B118}, 76
(1977).\\
5. R. Casalbuoni, Phys. Lett. {\bf B62}, 49 (1976); Nuovo Cim. {\bf
A33}, 389 (1976).\\
6. D.V. Volkov and A.I. Pashnev, Teor. Mat. Fiz. {\bf 44}, 331
(1980).\\
7. L. Brink and J.H. Schwarz, Phys. Lett. {\bf B100}, 310 (1981).\\
8. J.A. Azc\'arraga and J. Lukierski, Phys. Lett. {\bf B113}, 170
(1982); Phys. Rev. {\bf D28}, 1337 (1983)\\
9. W. Siegel, Class. Quant. Grav. {\bf 2}, L95 (1985); Nucl. Phys. {\bf
B263}, 93 (1985).\\
10. V.D. Gershun and V.I. Tkach, JETP Lett., {\bf 29}, 320 (1979).\\
11. R. Penrose and W. Rindler, {\it Spinors and Space-Time} (Cambridge
Univ. Press, Cambridge, 1986).\\
12. P.S. Howe, S. Penati, M. Pernici, and P. Townsend, Phys. Lett. {\bf
B215}, 255 (1988).\\
13. R. Marnelius and U. M\.{a}rtensson, Nucl. Phys. {\bf B335}, 395
(1991); Int. J. Mod. Phys. {\bf A6}, 807 (1991).\\
14. J. Frenkel, Z. Phys. {\bf 37}, 243 (1926).\\
15. K. Rafanelli, Phys. Rev. {\bf D30}, 1707 (1984).\\
16. J. Wess and J. Bagger, {\it Supersymmetry and Supergravity}
(Princeton Univ. Press., Princeton, 1983).\\
17. H.M. Farkas and I. Kra, {\it Riemann Surfaces} (Springer-Verlag,
N.Y. 1980).\\
18. P.A.M. Dirac, {\it Lectures on Quantum Mechanics} (Yeshiva Univ.
Press, N.Y. 1964).\\
19. V. Bargmann and E.P. Wigner, Proc. Nat. Acad. Sci. {\bf 34}, 211
(1947).\\
20. S.M. Kuzenko, S.L. Lyakhovich, and A.Yu. Segal, in preparation.\\
21. S.M. Kuzenko, S.L. Lyakhovich, A.Yu. Segal, and A.A. Sharapov, in
preparation.\\
22. A.S. Galperin, P.S. Howe, and K.S. Stelle, Nucl. Phys. {\bf B368},
248 (1992).\\
23. D.P. Sorokin, V.I. Tkach, and D.V. Volkov, Mod. Phys. Lett. {\bf
A4}, 901 (1989); D.P. Sorokin, V.I. Tkach, D.V. Volkov, and A.A.
Zheltukhin, Phys. Lett. {\bf B216}, 302 (1989).\\
24. E.S. Fradkin and M.A. Vasiliev, Ann. Phys. (N.Y.) {\bf 177}, 63
(1987); M.A. Vasiliev, Fortschr. Phys. {\bf 36}, 33 (1988).\\
25. E.S. Fradkin and M.A. Vasiliev, Nucl. Phys. {\bf B291}, 141 (1987).

\end{document}